\documentclass[11pt,letterpaper]{article}
\pdfoutput=1
\usepackage{jheppub}
\usepackage{color}
\usepackage{graphicx}
\usepackage{wrapfig}
\usepackage{epstopdf}

\usepackage{verbatim}
\usepackage{amsmath}
\usepackage{amssymb}
\usepackage{subfig}
\usepackage{url}
\usepackage{bbold}
\usepackage{xspace}
 \usepackage{setspace}

\usepackage{floatrow}
\newfloatcommand{capbtabbox}{table}[][\FBwidth]

\usepackage{slashed}
\usepackage{verbatim}

\def\beq{\begin{eqnarray}}
\def\eeq{\end{eqnarray}}
\def\bea{\begin{eqnarray*}}
\def\eea{\end{eqnarray*}}

\def\centeron#1#2{{\setbox0=\hbox{#1}\setbox1=\hbox{#2}\ifdim
\wd1>\wd0\kern.5\wd1\kern-.5\wd0\fi
\copy0\kern-.5\wd0\kern-.5\wd1\copy1\ifdim\wd0>\wd1
\kern.5\wd0\kern-.5\wd1\fi}}
\def\ltap{\;\centeron{\raise.35ex\hbox{$<$}}{\lower.65ex\hbox{$\sim$}}\;}
\def\gtap{\;\centeron{\raise.35ex\hbox{$>$}}{\lower.65ex\hbox{$\sim$}}\;}

\newcommand{\newc}{\newcommand}
\newc{\qbar}{{\overline q}}
\newc{\Kahler}{Kahler }
\newc{\deltaGS}{\delta_{\rm GS}}

\begin{document}

\title{Challenges for the Nelson-Barr Mechanism}

\author[a]{Michael Dine}
\author[b,a]{and Patrick Draper}

\affiliation[a]{Santa Cruz Institute for Particle Physics and Department of Physics, University of California, Santa Cruz, CA 93106, USA}
\affiliation[b]{Department of Physics, University of California, Santa Barbara, CA 93106, USA}

\emailAdd{mdine@ucsc.edu}
\emailAdd{pidraper@physics.ucsb.edu}

\date{\today}

\abstract{The axion and $m_u=0$ solutions to the strong CP problem have been subject to the most careful scrutiny and critique.  Basic theoretical issues include hierarchy and fine-tuning problems, quality and genericity of symmetries, and compatibility with solutions to the electroweak hierarchy problem. We study the similar set of challenges for solutions to strong CP based on spontaneous CP violation and the Nelson-Barr mechanism. Some of our observations have appeared in the literature previously, and others are new; our purpose is to collect and analyze the issues as a whole and provide an assessment of the most plausible settings for the Nelson-Barr solution.
}


\arxivnumber{1506.05433}

\preprint{}

\maketitle

\section{Introduction}
There are various naturalness problems of the Standard Model (SM), including the cosmological constant problem, the hierarchy problem, the hierarchies in the quark and lepton mass matrices, and the strong CP problem. Of these, the last is special.  Even modest changes in the cosmological constant  would drastically alter the world around us.  Similarly, the values of the weak scale and the light quark and lepton masses
play critical roles in a range of phenomena.   But if the CP-violating parameter $\bar \theta$ were, say, $10^{-3}$, there would be no
appreciable change in nuclear physics.  

Theorists may put forward complicated explanations for the smallness of $\bar \theta$, with many additional degrees of freedom, complicated symmetries, and some amount of fine tuning, but this activity is not particularly satisfying.
More compelling would be a theory in which the the smallness of $\bar \theta$ 
emerged as an accidental consequence of other structure in a physical theory:  an explanation of flavor or dark matter, for example. 
We will refer to this (presently hypothetical) phenomenon as {\it incidental CP conservation}.

Most attention has focussed on three solutions to the strong CP problem:  the possibility of a massless up quark, the Peccei-Quinn (PQ) solution with its associated axion~\cite{Peccei:1977hh,Peccei:1977ur},  and spontaneous CP or P violation with a protection mechanism for $\bar\theta$~\cite{Beg:1978mt,Georgi:1978xz,Mohapatra:1978fy,Segre:1979dx,Barr:1979as,nelsoncp,barrcp,barrcp2,Babu:1989rb,Barr:1991qx,Kuchimanchi:2010xs}.   The first two solutions require that the theory possess an approximate $U(1)$ symmetry, the violation of which is primarily due to the QCD anomaly. If the symmetry is not spontaneously broken
at scales above the QCD scale, there must be one or more very light quarks.    
This is usually stated as the requirement that the $u$ quark mass vanishes, but the more precise statement is that
at scales beyond a few GeV, ${m_u \over m_d} < 10^{-10}$.  Apart from any theoretical issues, the possibility of a massless quark is strongly disfavored by lattice calculations~\cite{Aoki:2013ldr}.
If the chiral symmetry is nonlinearly realized, there is a light axion~\cite{Peccei:1977hh,Peccei:1977ur}.  The potential for this axion determines $\bar \theta$.  

The third proposed solution is that CP or P is spontaneously broken and $\bar \theta$ is protected by extra structure~\cite{Beg:1978mt,Georgi:1978xz,Mohapatra:1978fy,Segre:1979dx,Barr:1979as}, the most common example of which is the Nelson-Barr (NB) mechanism~\cite{nelsoncp,barrcp,barrcp2} in the case of spontaneous CP violation.\footnote{In the interesting alternative case of spontaneous parity violation, models and their criteria for success were discussed in~\cite{Babu:1989rb,Barr:1991qx,Kuchimanchi:2010xs}.  Another mechanism in the case of spontaneous CP violation,  distinct from NB, involves the introduction particular ``shaping symmetries" in the underlying flavor structure~\cite{Antusch:2013rla}.}  Since the underlying theory is CP-conserving, the ``bare" $\bar \theta$
parameter vanishes.  CP must then be spontaneously broken in a way that ensures a small effective $\bar \theta$ while allowing an order one phase in the
CKM matrix (and a mechanism for baryogenesis)~\cite{Beg:1978mt,Georgi:1978xz,Mohapatra:1978fy,Segre:1979dx,Barr:1979as,nelsoncp,barrcp,barrcp2}.  The NB proposal is striking in that it seeks to solve the strong CP problem with no low energy consequence, unlike the axion and $m_u=0$ solutions.   On the other hand, in this paper, we will see some relations between these proposals.\footnote{Other solutions~\cite{Rubakov:1997vp,hook,hillerschmaltz} possess close similarities to the solutions with approximate U(1)s~\cite{Rubakov:1997vp,hook} or NB~\cite{hillerschmaltz}.}

Setting aside the possibility that $m_u=0$ leaves the PQ and NB proposals.  As currently implemented
in an array of models, neither is completely satisfactory from a theoretical point of view; certainly neither is obviously incidental in the sense
defined above.  For the PQ solution, the theoretical problems have been extensively discussed, and we will review some of the issues.  The primary focus of this paper will be the challenges to obtaining a plausible implementation
of the NB solution.  In both PQ and NB, the inadequacies of current proposals concern the structure of the
microscopic, ultraviolet theory and particularly the complexity and plausibility of the structures necessary for an effective solution.  
\begin{enumerate}
\item  The principal difficulty with the axion mechanism is that the PQ symmetry needs to be of very high quality. If
this symmetry is an accident, it must be a remarkably good one.  If the symmetry and its breaking are described by a conventional effective
field theory, the required quality can be achieved with a $Z_N$ symmetry, but requires $ N \ge 11$ or so. This is hardly a compelling explanation for the
smallness of an inconsequential parameter of the Standard Model.\footnote{In \cite{Carpenter:2009zs}, the possibility that $N$ is large in order to account for dark matter was considered.  It was shown that dark matter can account for a large value of $N$, but not large enough to solve the strong CP problem.}  In string theory, the situation for light axions appears better, but a solution
in this framework requires assumptions about the stabilization of moduli which, while perhaps imaginable, at least at present are impossible
to verify.  In the string framework, one must also hypothesize an unconventional cosmology and typically some tuning of initial conditions, unless the axion decay constant is surprisingly small.
\item  
As we will elaborate in this paper, the NB mechanism is 
generically on even weaker theoretical ground.
If the implementation is not massively fine-tuned, it requires strong dynamics or supersymmetry (though not necessarily at scales of order a few TeV). Strong dynamics are insufficient to protect small $\bar\theta$ in the simplest models, and supersymmetric models require gauge mediation ($m_{3/2} \ll$ splittings in supermultiplets). In addition,
new discrete or gauge symmetries and strong coincidences of scales are necessary, as well as
a number of degrees of freedom beyond those required by supersymmetry.  
\end{enumerate}

Instead of such speculative exercises, one can hope for an experimental resolution.  The discovery of an axion would, needless to say, answer the question.  However, a large part of the axion parameter space is currently inaccessible. For the NB solution, there is no similar ``smoking gun."  While we will argue that gauge mediation is a requirement, the scale need not be particularly low.

This paper is organized as follows.
In Section \ref{nelsonbarrmodels}, we review the basic structure of the fermionic sector of NB models. In Section \ref{nonsusy}, we discuss non-supersymmetric models. 
If such models contain fundamental scalars, one would expect the scale of CP violation to be high in order to limit the fine-tuning.  However,
constraints imposed by dangerous higher-dimension couplings require a low scale of CP violation, implying enormous fine tuning.
Although compositeness can explain the required hierarchy, we argue that the simplest models typically fail to retain the necessary NB structure. Setting the fine-tuning issue aside, we discuss the sorts of symmetries which might ensure  vanishing $\bar \theta$ at tree level, and discuss the dangerous radiative corrections to $\bar \theta$ that can arise at one and two loop order.
In Section \ref{susyCP} we turn to supersymmetry.  In theories for which supersymmetry is broken well below some
``fundamental" ultraviolet scale (perhaps the Planck, string, or compactification scale),  we can pose more sharply the question of what it
means for the bare $\theta$ to vanish.
We argue that in practice there is a heavy axion, and thus a sense in which the supersymmetric NB and PQ models can be considered
as different limiting cases of axion models.  We discuss
how the expectation value of this axion might be fixed and constraints on couplings of the axion to possible CP-violating sectors.  We also note that
very simple landscape considerations suggest that vanishing of the ``bare $\theta$" in such frameworks is {\it extremely} rare, and these
is no obvious anthropic selection effect one might invoke. Finally, we discuss the spontaneous breaking of $CP$ and SUSY and the radiative corrections to $\bar\theta$ in supersymmetric models with gravity and gauge mediation.   In gravity mediation, corrections are typically large and spoil the NB solution. In gauge mediation, the corrections can be smaller, but there are upper bounds on the ratio of the susy-breaking scale to the scale of CP violation. In Section \ref{conclusions} we summarize and conclude.  

\section{The Essence of the Nelson-Barr Mechanism}
\label{nelsonbarrmodels}

The main challenge in solving the strong CP problem with spontaneous CP violation is to understand why
\beq
{\rm Arg ~det ~m_q} < 10^{-10},
\eeq
while there is a large phase in the CKM matrix. Nelson~\cite{nelsoncp} and Barr~\cite{barrcp,barrcp2} obtained the first simple, phenomenologically viable models which
achieve this and elucidated the general properties of renormalizable Lagrangians that can exhibit ${\rm Arg ~det ~m_q}=0$ at tree level. 

A model with minimal field and symmetry content was obtained by Bento, Branco, and Parada (BBP)~\cite{brancoetal}, and serves as a useful starting point for understanding the properties of the NB mechanism. The BBP model introduces additional charge $\pm 1/3$ $SU(2)$ singlet quarks $q$, $\bar q$, as well as a set of complex fields $\eta_a$ neutral under the SM (we will comment on real fields later).  
The down-type quark mass terms in the BBP model are given by
\beq
{\cal L} = \mu \bar q q + a_{af} \eta_a \bar d_{\bar f} q+ y_{f \bar f} HQ_f \bar d_{\bar f}+ \dots\;.
\label{basicnelsonbarr}
\eeq 
The $\eta_a$ are assumed to have vevs with relative phases, breaking CP.\footnote{In fact, in the original BBP model~\cite{brancoetal}, only a single complex field is introduced with Yukawa couplings $(a_f \eta + a_f^\prime \eta^*)\bar d_{\bar f} q$. This structure is sufficient as long as $a_f$ and $a_f^\prime$ are nonzero, $a_f\neq a_f^\prime$, and a required discrete symmetry under which $\eta$, $q$, and $\bar q$ transform is a $Z_2$ instead of a more general $Z_N$. We consider the form of Eq.~(\ref{basicnelsonbarr}), with multiple $\eta_a$ and vanishing $a_{fa}^\prime$, anticipating possible $Z_N$ symmetries as well as the extension of the BBP model to supersymmetry.}

At tree level, the Lagrangian in~(\ref{basicnelsonbarr}) automatically gives ${\rm Arg~det~m_q} = 0$ for the quark masses. However, it is not the most general renormalizable Lagrangian allowed by the symmetries of the SM. Couplings of the form  $\eta_a q \bar q$ and $HQ\bar q$ must be forbidden. Similarly, we might like $\mu$ to be the expectation value of a CP-conserving field, which constrains its interactions with the $\eta_a$. Discrete symmetries can provide the necessary structure, and we return to this issue in the next section.

The CKM phase in the SM is generated by integrating out the heavy flavor from~(\ref{basicnelsonbarr}).  Defining the
$4\times 4$ quark mass matrix as:
\beq
{\cal M} = \left ( \begin{matrix} \mu & B  \cr 0 & m_d \end{matrix} \right ) ;~m_d \equiv yv;~B_f = a_{af} \eta_a\;,
\eeq
we need to diagonalize the matrix
\beq
{\cal M} {\cal M}^\dagger = \left ( \begin{matrix} \mu^2 + BB^\dagger & B m_d^T \cr
m_d B^\dagger & m_d m_d^T \end{matrix} \right )\;.
\eeq
If the left hand corner of this matrix is larger than the other entries, we can
integrate out the heavy state, leaving the $3 \times 3$ SM mass matrix:
\beq
\left (  (m_dm_d^T )_{ij} - {(m_d)_{ik} B^\dagger_k B_\ell (m_d^T)_{\ell j} \over \mu^2 + B_fB_f^\dagger} \right ) \;.
\label{eq:SMmatrixBBP}
\eeq
The diagonalizing matrix is the CKM matrix.
Note that this procedure is correct only in the limit
$\mu^2 + \vert B_f \vert^2 \gg m_d^2$; otherwise, the CKM matrix is not unitary.  

Obtaining a large CKM phase strongly constrains the parameters. If there is only one non-vanishing $B_f$,
or if each $B_f$ has the same phase, or if $\mu \gg \vert B_f \vert$, then the CKM matrix is real.  
However, if there are two distinct, non-vanishing $B_f$ of comparable magnitude and with a large relative phase, and $\mu\lesssim \vert B_f \vert$, there is a non-trivial
phase.
For example, if $B = \left ( 0,~  b,~  c \right )$, a phase of order ${\rm Im(b/c)}$ enters the CKM matrix. We see that a rather close coincidence of scales is required between the real and imaginary parts of different fields.  The severe challenges for non-susy NB theories will be discussed
in the next section.

\section{Nonsupersymmetric Nelson-Barr Models}
\label{nonsusy}

In this section we consider nonsupersymmetric Nelson-Barr models. We begin with a survey of the basic issues and challenges confronting such models already at tree level, and then elaborate on two of the issues that arise when radiative corrections are included.

\subsection{Basic Challenges}

Without supersymmetry, it is a simple matter to construct models of spontaneous CP violation.  We can, for example, introduce two real fields, $\sigma$ and $\pi$, the first CP-even and the second CP-odd, with appropriate NB-type couplings to fermions and a potential that leads to a vev for each. Likewise with complex fields it is not difficult to spontaneously break CP, if there is sufficient freedom in the specification of the scalar potential (for a principled discussion of necessary and sufficient conditions, see~\cite{Haber:2012np}.

However, NB models, to be viable, must confront several theoretical challenges:
\begin{enumerate}
\item  Further symmetries are necessary to enforce the necessary structure of the mass matrix, even at the renormalizable level. In the BBP model discussed in the previous section, since $\mu\lesssim|\langle \eta_a \rangle|$, it is necessary suppress or forbid dimension-4 couplings of the form $\eta_a q \bar q$.  Likewise we must suppress $HQ\bar q$. One possibility is to allow the new scalars and fermions to transform under a $Z_N$ symmetry (if $N>2$, then the scalars must be complex, as in the model discussed above):
\beq
\eta_a \rightarrow e^{2 \pi i k \over N}\eta_a\;,~~~q_f \rightarrow e^{-{2 \pi i k \over N}} q_f\;, ~~~\bar q_f \rightarrow e^{{2 \pi i k \over N}} \bar q_f\;.
\label{eq:ZN}
\eeq
With other fields neutral, we obtain a Lagrangian of the desired form.  It is not difficult to write down models which spontaneously break both CP and the $Z_N$.   We will discuss possible gauge symmetries when we consider supersymmetry in the next section.
\item  The scale of spontaneous CP breaking $m_{CP}$ should be low compared to the cutoff $\Lambda$.  Dimension-5 operators such as
\beq
\eta_a^* \eta_b \bar q q\;,~~~{\eta_a} HQ\bar q
\label{eq:HDO}
\eeq
for example, can induce $\bar\theta$ of order $(m_{CP} / \Lambda)$. Note that the $Z_N$ symmetry defined in Eq.~(\ref{eq:ZN}) (or possible
$U(1)$ symmetries) does not help to suppress higher-dimension operators like~(\ref{eq:HDO}). Without further symmetries or fine-tuning, even if the cutoff is $\Lambda=M_p$, suppression of such operators requires
\begin{align}
 m_{CP}\lesssim 10^{8}~{\rm GeV}\;.  
\end{align}
\item  As in any non-supersymmetric or non-composite model, light scalars are fine-tuned.  Here we require at least
two such scalars at a scale $m_{CP}\ll M_p$, and the fine-tuning of {\it each} of these masses is much worse than just fine-tuning $\bar\theta$ by itself.  {\it It is difficult to make sense of NB models outside of a broader framework in which $m_{CP}/M_p$ is naturally small.} 
\item  As we have seen in the previous section, to obtain a substantial CKM angle, it is critical that the expectation values of different CP-odd and CP-even fields (times suitable couplings) coincide to better than an order of magnitude.  
\item   We might want to account for $\mu$ dynamically, i.e. through the expectation value of a fundamental or composite
field $S$.  Additional symmetries need to be introduced to avoid inducing phases in $S$ from couplings of $S$ to the $\eta_a$.
\item Even when it vanishes at tree-level, $\bar \theta$ is often generated radiatively at the scale $m_{CP}$.  
\end{enumerate}

Loop effects are particularly
problematic.  They cannot be suppressed simply by additional (bosonic) symmetries or by lowering the scale of CP violation.
These corrections will be the subject of the next section.

\subsection{Radiative Corrections to $\theta$ in Non-Supersymmetric Theories}

Even if one closes one's eyes to fine tunings, and one is willing to accept a low scale for CP violation, loop corrections
are quite problematic in NB models.  Threshold corrections to $\bar\theta$ have to be considered on a model-by-model basis, but certain operators are typically problematic. BBP studied $\bar\theta$ at one loop in~\cite{brancoetal}. Below, we review and reinterpret their result, and observe further problematic contributions at two loop order. We will see that the one loop sensitivity of $m_{CP}$ to the UV cutoff requires us to add structure, such as supersymmetry or a dynamical origin for the scalars, and then to consider all of the other issues in that larger framework.  In the subsequent section we discuss composite models and see that while the fine-tuning of $m_{CP}$ can be resolved, simple cases will either have difficulty maintaining $\bar\theta=0$ at tree level, or will have one loop corrections to $\bar\theta$ similar to non-composite models. This will lead us to consider NB in the supersymmetric context.

\begin{figure}[t]
\centering
\includegraphics[width=0.5\textwidth,trim={4cm 15cm 4cm 2cm},clip]{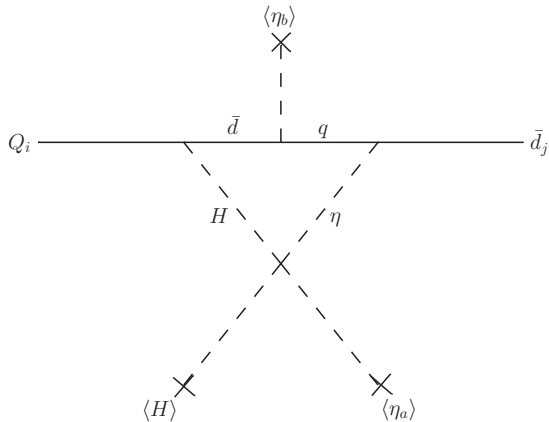} 
\caption{Example threshold correction to Arg det $m_d$.}
\label{fig:quarkphaseBBP}
\end{figure}

In the BBP model, dangerous contributions to $\bar\theta$ arise at one loop from the Higgs portal operators
\beq
(\gamma_{ij}\eta_i^\dagger \eta_j + \lambda_{ij} \eta_i \eta_j + cc) H^\dagger H \;.
\label{eq:HP}
\eeq
$\lambda_{ij}$ can be forbidden by a $Z_N$ symmetry with $N>2$, so we consider the effects of $\gamma_{ij}$. 
Unless the $\gamma$s are very small, these couplings make a large contribution to the Higgs mass. In the context of a solution to the $m_{CP}$ hierarchy problem, there might or might not be a principled reason why the couplings are small, but {\it a priori} they indicate only another contribution of many to the tuning of $m_H^2$.  At one loop, the diagram of Fig.~\ref{fig:quarkphaseBBP} gives a complex correction to the SM down-type Yukawa coupling, contributing to a shift in $\bar\theta$ of order
\begin{align}
\Delta \bar\theta \simeq {\rm Im~Tr~}y^{-1} \Delta y \simeq \frac{\eta_a a_{af} a_{bf}\gamma_{bc}\eta_c^*}{16\pi^2 m_{CP}^2}\;.
\end{align}
Adequately suppressing $\bar\theta$ requires the $a$ and/or $\gamma$ couplings to be small.

The authors of~\cite{brancoetal} took the viewpoint that whatever solves the SM hierarchy problem might suppress
the portal couplings.  Such suppressions can occur in supersymmetric or composite theories (both of which solve the $m^2_{CP}$ hierarchy problem, but not necessarily the full $m_H^2$ one).  These theories involve significant extra structure beyond the minimal BBP model, and the radiative corrections to $\bar\theta$ must be considered in the full theories.  Without supersymmetry or extra dynamics, the Higgs mass is simply tuned, and small $\theta$ is problematic.

At two loop order, there are additional contributions which must be suppressed.  In particular, insertions of the operator
\beq
{\cal L}_{\eta^4} = \gamma_{ijkl} \eta_i \eta_j \eta_k^* \eta_\ell^*
\eeq
can contribute phases to the operators $\mu \bar q q$ and $Q H \bar d$.
\begin{figure}[t]
\centering
\includegraphics[width=0.5\textwidth,trim={4cm 20cm 4cm 2cm},clip]{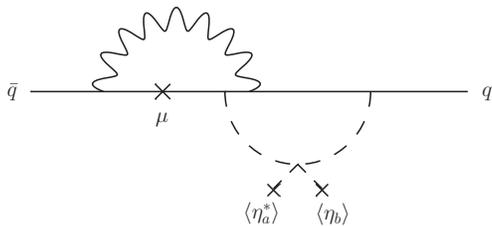} 
\caption{Example two-loop contribution to the phase of $\mu$.}
\label{fig:muphaseBBP}
\end{figure}
The relevant Feynman diagrams contain a loop of gauge bosons and an $\eta$ loop, with
insertions of ${\cal L}_{\eta^4}$; an example is given in Fig.~\ref{fig:muphaseBBP} (this contribution is similar to the ``dead duck" graph noted in~\cite{nelsoncp}).  The contribution to $\bar\theta$ is of order
\begin{align}
\Delta\bar\theta\simeq\frac{g^2 a_{af}a_{cf}\eta_b^*\eta_d\gamma_{abcd}}{(16\pi^2)^2m_{CP}^2}
\end{align} Again, unless the couplings are surprisingly small, the correction is several orders of magnitude to large. In the supersymmetric case, we will see that these contributions can be suppressed, but new issues will arise.

\subsection{Models with Strong Dynamics}
\label{strongdynamics}

The low scale of CP violation may be protected by strong dynamics. For example, the CP-odd scalars could be pseudogoldstone mesons $\Pi$ of an SU(N) gauge theory in which condensates spontaneously break approximate chiral flavor symmetries,
\begin{align}
\langle\bar\psi_i\psi_j\rangle=B f_\Pi^2 {\rm exp}(i\Pi^at^a/f_\Pi)\;,
\end{align}
in analogy with the pions of QCD. The $\Pi$ fields can obtain nonzero vevs naturally from a particular pattern of chiral symmetry breaking (as in, e.g., Dashen's model~\cite{Dashen:1970et}). In this case, BBP-type couplings to the Standard Model and the $q$,$\bar q$ messengers (assumed for now to be fundamental fermions) might arise from higher-dimensional operators of the form
\begin{align}
\frac{1}{\Lambda^2}\kappa^f_{ij}\bar\psi_i\psi_j\bar d_f q/\Lambda^2\;\rightarrow\; B\frac{f_\Pi^2}{\Lambda^2}{\rm Tr}\left[\kappa^f e^{i\Pi^at^a/f_\Pi}\right]\bar d_f q+\dots\;.
\label{eq:compositeop}
\end{align}
If the hierarchy between the scale of the gauge theory $\sim f_\Pi$ and the UV cutoff $\Lambda$ is large, the effective couplings $a_{af}$ in Eq.~(\ref{basicnelsonbarr}) may be very small, and the effective scale of CP violation much smaller than $f_\Pi$. We can see from the form of Eq.~(\ref{eq:SMmatrixBBP}) that the CKM phase can still be large if $\mu$ is sufficiently small. Furthermore, the one loop BBP radiative correction -- generated here by couplings of the form $H^\dagger H \bar \psi \psi /\Lambda$ -- is suppressed when the effective $a_{af}$ couplings are small.

Unlike in the fundamental scalar case, however, it is difficult to implement discrete symmetries needed to keep $\mu$ real. Permitting~(\ref{eq:compositeop}) while forbidding the similar 4-fermi operator $\bar \psi \psi \bar q q$ requires the discrete symmetry to act chirally on $\psi$,$\bar\psi$ (and, for example, on $q$,$\bar q$), but explicit chiral symmetry breaking is necessary to generate the spontaneous CPV potential when the CP-odd scalars are pseudogoldstones.  This breaking might be soft, as in a set of masses $m$ for the $\psi$,$\bar\psi$, and thus the coefficient of $\bar \psi \psi \bar q q/\Lambda^2$ might be suppressed by $m/\Lambda$. But if $m$ is not too different from $f_\Pi$, then $f_\Pi/\Lambda$ must be less than $10^{-10}$, resulting in an unacceptably low value for $m_{CP}$.

It is even more difficult to understand the NB structure and the reality of the effective $\mu$ if the messenger fields $q$,$\bar q$ are baryons of the gauge theory. In this case the baryon mass is expected to arise principally from spontaneous chiral symmetry breaking, which by construction breaks CP.

We stress that it is not impossible to build NB-type models with strong dynamics, but it requires more complicated structures. A minimal example was constructed in Ref.~\cite{Vecchi:2014hpa}, consisting of a BBP-type model in which the $\bar \psi \psi \bar q q$ operator is forbidden by a gauged subgroup of the chiral flavor symmetry. This symmetry might also be discrete. The Dashen mass terms are forbidden by the symmetry, but the potential can still break CP with suitable dimension-6 operators $(\bar \psi \psi)^2$.  Ref.~\cite{Vecchi:2014hpa} also showed that models with acceptably small radiative corrections to $\bar\theta$ could be distinguished by the flavor transformation properties of the CPV spurions present in the low-energy theory. BBP-type models with generic couplings possess CPV spurions in the infrared in both the fundamental and anti-fundamental representations of $SU(3)_d$, and as such they fail the criteria of~\cite{Vecchi:2014hpa}. This is reflected in the large one loop correction to $\bar\theta$. However, when the couplings $a_{af}$ are small, as can arise in strongly-coupled models as discussed above, the low-energy theory contains only an $SU(3)_d$-fundamental spurion and the criteria for small corrections to $\bar\theta$ are met.

\section{CP in Supersymmetric Theories: Axions, Moduli, and $\theta$ at Tree Level}
\label{susyCP}

Supersymmetry, with SUSY breaking at scales well below
the scale of CP violation, can significantly ameliorate the Nelson-Barr fine-tuning problem.  In addition, SUSY can forbid some of the problematic
higher-dimension operators and quantum corrections to $\bar \theta$ encountered in the non-SUSY case. In this section,
we consider supersymmetric Nelson-Barr models and their symmetries.  We first review some of the problematic aspects
of the Peccei-Quinn solution of the strong CP problem and their possible resolution.  Then we consider more carefully the underlying premise that CP can naturally be a good symmetry, and as a result that the
bare $\bar\theta$ vanishes. In both cases the questions are ultraviolet-sensitive and the resolutions depend on the structure of the microscopic theory.  In particular, if there is an underlying
landscape, small bare $\bar \theta$ is implausible. 

We first review some aspects of the axion solution, with and without supersymmetry.
The most challenging aspect of the Peccei-Quinn solution of the strong CP problem is understanding why the global symmetry is so good.
Global symmetries should arise only as accidents of gauge symmetry and the structure of low dimension
terms in an effective action.
It was quickly recognized that this is a challenge for the PQ mechanism~\cite{kamionkowski}.  
From a PQ-violating potential $V_{pqv}$, we can define an axion quality factor,
\beq
Q_a \equiv {f_a {\partial V_{pqv}(a) \over \partial a} \over m_\pi^2 f_\pi^2 }\;.
\eeq
Solving the strong CP problem requires
\beq
Q_a < 10^{-10}\;.
\eeq
In a conventional effective field theory analysis (i.e. finite number of degrees of freedom above
$f_a$), small $Q_a$ is highly non-generic.  If the axion arises as the phase of a field $\Phi$,
\beq
\langle \Phi \rangle = f_a e^{ia/f_a}\;,
\eeq
symmetry violating operators like 
\beq
{\Phi^{n+ 4} \over M_p^n}
\eeq
spoil the PQ mechanism even for $f_a = 10^{11}$ GeV unless $n>7$.
Such suppression can be obtained with a discrete $Z_N$ symmetry, with $N\geq 11$, but such a model appears contrived.

Witten pointed out early on that string theory provides a possible resolution to the problem of the quality of the PQ symmetry~\cite{wittenaxion}.
This is most easily understood in the framework of supersymmetry.  Typically
string models possess moduli, $\Phi$, whose imaginary component obeys a discrete shift
symmetry:
\beq
\Phi = x + i a;~~a \rightarrow a + 2 \pi
\label{eq:moduli}
\eeq
This symmetry guarantees that any superpotential is a function of $e^{-\Phi}$ at large $x$.
Here $x$ might be $8 \pi^2 \over g^2$, for some gauge coupling $g$.

In this setting, the primary question is why the theory sits in an asymptotic region of the moduli space
where $e^{-x}$ is very small.  It is consistent at least
with the fact that the observed gauge couplings are small, but a detailed connection is not possible at present, much less reliable computations~\cite{dinefestucciaaxions}.

We turn now to theories where CP is a symmetry of the microscopic dynamics.
Here we can make a connection with string axions discussed above. In known string theories, CP is a good symmetry~\cite{stromingerwittencp,kaplannelsoncp,dineleighcp}.  For typical string compactifications, this statement
means that there is a subspace of the moduli space on which CP is conserved, and CP is spontaneously broken on the rest.   In supersymmetric
theories, the moduli fields include both a CP-even and a CP-odd scalar, as in Eq.~(\ref{eq:moduli}), and we will refer to them as saxions $x_i$ and axions $a_i$, respectively.  We can define $a_i =0$
as the CP conserving point. CP is spontaneously broken
if some of these axions are stabilized at $a_i \ne 0$.  Generally one or moduli couple to each of the gauge groups in the classical theory,
providing candidate axions.  The question of whether there is a non-zero $\theta$ is then a question of whether the
relevant axions are heavy and fixed at CP conserving points.

If the moduli are stabilized supersymmetrically, the CP-even and CP-odd states are fixed together.  Suppose that we have a single modulus, 
with
\beq
W = -\alpha e^{-\Phi/b} + W_0;~~K = -\log(\Phi + \Phi^\dagger).
\eeq
with $W_0$ small, as in the KKLT scenario\cite{kklt}.   
Then
\beq
\Phi \approx b \log (W_0/\alpha)\;.
\eeq
Provided $W_0$ and $\alpha$ are real, $\Phi$ is real.  If $\Phi$ couples to the QCD gauge fields as $\Phi W_\alpha^2$,
it generates no tree-level contribution to $\theta$.  Plausibly, if $W_0$ is large, CP remains unbroken, and $\Phi$ is very heavy.

Should $W_0$ be real? If we assume $W_0$ results from CP-conserving dynamics, it is automatically real.  On the other hand,
flux landscapes provide a model where complex $W_0$ appears more likely.  In such cases $W_0$ is the sum of many contributions associated
with many different fluxes, of which we expect about half  to be CP-even and half to be CP-odd.  CP preservation amounts to requiring half of the
of the fluxes to vanish.  In other words, given $10^{500}$ states, only $10^{250}$ conserve CP and have vanishing $W_0$, and correspondingly CP-conservation appears very non-generic.  Moreover, as noted earlier, it is hard to see what might select for small $\theta$.  However, absent a sharp UV prediction for $W_0$, we can simply take its reality as a requirement of the NB setup. 

We can ask what may happen when we introduce a sector
in which CP is spontaneously broken with characteristic scale $\mu$. If this sector does not break supersymmetry, 
we might expect additional, CP-violating terms in the superpotential of order $\mu^3e^{-S}$.  These terms will shift the minimum of
the axion field, but their contribution is suppressed if $b$ is large.  If, for example, $e^{-S} < 10^{-15}$ and $b=5$, then $\theta < 10^{-12}$.  Alternatively, if $b=1$, the contribution to $\theta$ is suppressed by at least ten order orders of magnitude provided the scale $\mu$ is at least three orders of magnitude 
below $M_p$. In non-supersymmetric models (e.g. cases where the scale of SUSY-breaking is $\gg \mu$) with axions, one would expect the difficulties to be at least as severe; it is not clear in such
contexts that terms violating the Peccei-Quinn symmetry must be exponentially small.

The assumption that $W_0$ is real constrains a combination of the supersymmetry breaking and CP violating scales.  In particular, we might expect CP violation to generation a complex term in the superpotential, $W_0 \sim \mu_{CP}^3$.  If there is no suppression
of the phase, the requirement of cancellation of the cosmological constant yields the constraint:
\beq
\mu_{CP}^3 <  M_{3/2} M_p^2.
\eeq

\section{SUSY Nelson-Barr Models}
\label{susynb}

In this section, we 
 assume that any would-be axions are massive and fixed in a CP conserving manner.
We then ask what are the requirements on SUSY NB models required to account for a very small $\bar\theta$.
The Lagrangian of~(\ref{basicnelsonbarr}) naturally extends to a superpotential:
\beq
W = \mu \bar q q + \lambda_{af} \eta_a q \bar d_f + y_{f\bar f} H_d Q_f {\bar d}_f +\dots \;.
\label{nelsonbarrw}
\eeq
For the moment we continue to treat $\mu$ as a dimensionful constant.  While the absence of  undesirable renormalizable interactions like $\eta q\bar q$ and $H_dQ\bar q$ can be technically natural due to nonrenormalization theorems, they can be forbidden in a more principled way with, for example, discrete symmetries like~(\ref{eq:ZN}). Again a coincidence in scales among the $\eta_a$ vevs is required, as well as $\mu\lesssim |\lambda_{af} \eta_a|$.

As emphasized above, putting NB into a larger and more natural framework incurs new challenges. The prime example in SUSY models is that the $\eta_a$ must be sequestered from the supersymmetry breaking sector to avoid, e.g., giving phases to the gluino mass, among other problems~\cite{dinekaganleigh}.   We might expect
the SUSY breaking theory to exhibit either an exact (discrete) $R$ symmetry,
or at least approximate accidental one.  If there is an identifiable Goldstino field, $Z$ (assumed chiral), then couplings of the $\eta_a$ to $Z$ must be suppressed.  

Replacing $\mu$ by a dynamical field $S$ may be desirable and requires further symmetries. For example, it is critical to forbid renormalizable couplings between $S$
and the $\eta_a$.

\subsection{Breaking of CP and $Z_N$ in SUSY}
If CP is violated at or below the scale of supersymmetry breaking, the low-energy theory can be studied in the non-supersymmetric
framework of the previous section.  Therefore, we focus on CP violation at scales much higher than those of supersymmetry breaking.  We will not attempt to be
exhaustive, but we consider models that illustrate some of the challenges.  We consider two classes of models:
\begin{enumerate}
\item  Models in which the CP violating fields are fixed supersymmetrically.  Here there is a discrete set of vacua and all fields have mass
of order the scale of CP violation.
\item  Models in which the CP violating fields are fixed by SUSY breaking dynamics. We take the scale
of CP violation to be much larger than the scale of SUSY breaking; in this situation, CP is broken by fields in approximate flat directions.
\end{enumerate}

\subsubsection{CP broken by Supersymmetry-Conserving Dynamics}

To write a simple model that breaks $CP$ in isolated vacua, we introduce two fields $\eta_1$ and $\eta_2$, odd
under a $Z_2$ symmetry, and fields $X$ and $Y$ that are even.  We can also suppose an $R$ symmetry
(for simplicity we will take it to be continuous, but it can also be a discrete subgroup) under which $X$ and $Y$ have $R$ charge $2$ and the $\eta_i$ are neutral.  Then we can take the
superpotential to have the form, without loss of generality:
\beq
W = X \mu^2 + X(a \eta_1^2 + b \eta_1 \eta_2 + c \eta_2^2)
+ Y (a^\prime \eta_1^2 + b^\prime \eta_1 \eta_2 + c^\prime \eta_2^2).
\eeq
This superpotential typically has minima in which $\eta_1$ and $\eta_2$ have phases, breaking CP.
If $q$, $\bar q$ are both odd under the $Z_2$, with $R$ charge $1$, and $\bar d_f$ is even, with $R$ charge $1$, then we obtain the NB superpotential at the renormalizable level.  

There are a number of issues with models of this type.  In particular, if supersymmetry breaking is associated with a Goldstino superfield in a hidden sector, $Z$, these symmetries will not forbid $Z \eta_1 \eta_2$ couplings, leading
to CP violating phases in ordinary soft breaking terms.  $Z_N$ symmetries with larger $N$, while forbidding these couplings, require more structure in order to obtain a  superpotential that is both $Z_N$ invariant and spontaneously breaks CP (and $Z_N$).  

Another model for spontaneous CP violation has been presented in \cite{Barr:1996wx}.  In addition to a discrete symmetry, the model relies on a continuous global symmetry to suppress couplings which would induce $\theta$ at tree level.  If the $U(1)$ is replaced by a discrete subgroup, at least a $Z_3 \times Z_5$ symmetry is needed to suppress dangerous {\it renormalizable} operators.

\subsubsection{Theories with Flat Directions}

String theory constructions suggest another possibility which can lead rather naturally to the NB structure.
There are two elements.  First, string models often possess $U(1)$ symmetries beyond those of the Standard Model,
as well as additional fields, which can yield the required superpotential for the NB models.  Second, there are often
approximate flat directions in which CP-odd fields can obtain large expectation values.  Under suitable
conditions, these vevs may spontaneously break CP.

In particular, the gauge group $E_6$, familiar in Calabi-Yau compactifications of the heterotic string, suggests the possibility of two additional
$U(1)$s at some energy scale as well as several additional fields.  
In terms of $O(10) \times U(1) \subset E_6$, the $27$ of $E_6$ decomposes as
\beq
27 = 16_{-1/2} + 10_1 + 1_{-2}\;.
\eeq
We will treat the theory as if this symmetry is broken to the Standard Model $\times U(1) \times U(1)$.
Then we can list the fields and their charges under the two $U(1)$s:
\beq
Q, \bar e, \bar u = (-1/2,1);~~L, \bar d = (-1/2, -3);~~\bar q = (1,2); ~~~ q = (1,-2);~~~\eta = (-1/2,5);~~~
\eeq
$$~~~~
H = (1,2)~~~ \bar H=(1,-2)~~S=(-2,0).
$$
Note that the $\eta$ is essentially the right-handed neutrino of $O(10)$, while the $S$ is the field in $E_6$ outside of the $16$ or $10$.
$q,\bar q$, and $\ell, \bar \ell$ arise from the $10$ of $O(10)$.  Anomaly cancellation is readily satisfied by
including an additional $q$, $\bar q$, $\ell$, $\bar \ell$,
$\eta$, $S$ for each generation.  In addition, we assume that there is one additional $S$, $\bar S$ pair and one additional $\eta, \bar \eta$ pair (and allow the possibility of other incomplete multiplets, particularly for the Higgs field).

With these charge assignments, the most general cubic superpotential involving $S, \eta, q, \bar q$ and the ordinary matter fields
is precisely that of Eq.~(\ref{nelsonbarrw}).   Moreover, at the renormalizable level, the classical theory possesses flat directions with non-zero
$\eta_i, \bar \eta, S_i, \bar S$.  

The flat directions may be lifted by supersymmetry-breaking effects and dimension-5 operators.
If some of the soft masses in the flat directions are negative, some of the fields will receive large expectation values.  If there are quartic superpotential couplings, e.g.
${1 \over M_p}\eta_i \eta_j \bar \eta^2$ and ${1 \over M_p}S_i S_j\bar S^2$, then these expectation values are of the order
\beq
S^2,~\eta^2 \sim m_{susy} M_p.
\eeq
With several fields, there will typically be CP violating minima of the potential.

Many problematic higher-dimension operators are forbidden by holomorphy and the U(1)s. However, a surviving class of dimension-5 operators, $S_i \bar S \eta_j \bar \eta$, must be forbidden to avoid large phases in $S$.
These couplings {\it can} be forbidden by discrete symmetries.  One virtue of this type of model is that it is compatible
with the existence of a (discrete) $R$ symmetry, which can suppress couplings of the $\eta$ fields to
any would-be supersymmetry-breaking sector and possible messengers.

Another potential difficulty is the large size of the $\eta_i$ expectation values.  These are sufficiently large that, depending on the scale of supersymmetry breaking and the suppression scale,  they have the potential
to induce $\bar \theta$ through dimension-6 operators.

\subsection{Breaking of Supersymmetry}
\label{susybreaking}

We have already noted that supersymmetry breaking introduces new potential contributions to $\bar\theta$. 
Many of these contributions do not decouple, even as the supersymmetry breaking scale is taken arbitarily
large.  As a result,
a successful supersymmetric solution to strong CP requires
suppression of phases in the gluino mass, as well as a high degree of degeneracy, proportionality, and suppression of phases in squark masses and $A$-terms~\cite{dinekaganleigh}, regardless of the scale of supersymmetry breaking.

We distinguish two classes of models: those, like gravity-mediated models, where the soft breaking terms of the SM fields are of order $m_{3/2}$, and those, like gauge mediated models, where $m_{3/2}$ is parametrically smaller.

Consider first gravity-mediated models.  In these models, one general issue is $\langle W \rangle
\sim m_{3/2} M_p^2$.  If $\langle W\rangle$ is complex, this feeds into $\theta$ through phases, for example, at one loop in the gaugino mass (this is the familiar anomaly-mediated contribution).    In Section~\ref{susyCP}, we raised general questions about the reality of $\langle W \rangle$, and argued that in flux landscapes, at least, real $\langle W \rangle$ is unlikely.  More generally, apart from some sort of anthropic selection, no convincing mechanism has been put forward to account for the value of the cosmological constant.  So the failure of landscape models to account for small phases is troubling.

In gauge-mediated models, the situation can be significantly better.  Comparing the anomaly-mediated to the gauge-mediated gluino mass, we require
\begin{align}
{\alpha_s \over 4 \pi}{m_{3/2} \over m_{susy}} < 10^{-10}\;. 
\end{align}
 This constraint places a loose upper bound on the underlying scale of supersymmetry breaking if $W$ possesses an order one phase. 

In both gravity and gauge mediation, there may be other strong constraints, depending on the nature of supersymmetry breaking.  If supersymmetry is broken in a hidden sector through a gauge-singlet chiral field, $Z$, with 
$F_Z =f$, then any phase in $f$ can feed into soft breaking terms, yielding phases for the gluino, for example, as well as squark mass matrices.  These, in turn, contribute to $\theta$.  In the models we have studied, these might arise from couplings such as
\beq
W_{\eta-Z} = \lambda \eta_i \eta_j Z
\eeq
at dimension three in $W$, or even through terms of dimension $2$.  Such undesirable terms can be forbidden if $Z$ is charged under some symmetry (as in some models of dynamical supersymmetry breaking), or by combinations of continuous and discrete symmetries in the models of CP breaking by pseudomoduli of the sort discussed in the previous section.   For example, couplings of combinations like $\eta_i \bar \eta$ to
$Z$ can be forbidden by $R$ symmetries.  In the models with discrete vacua, this problem is more challenging.  In
gauge-mediated models, it is also necessary to forbid couplings of the $\eta$ fields to messengers.  This can again arise from the $R$ symmetries consistent with the flat direction models.

If non-renormalizable terms coupling CP-breaking fields to $Z$ are permitted by symmetries, these will constrain the
scale of CP violation.  Certain Kahler potential terms are difficult to suppress by symmetries.
However, one can contemplate higher scales of CP violation than in the non-supersymmetric
case.

Overall, then, both in gravity and gauge mediation, it appears possible to avoid dangerous new sources of phases at tree level, without large
arrays of new fields or excessively 
complicated new symmetry structures. Gravity mediation requires stronger constraints on the reality of $W$.

\subsection{Loop Corrections in Supersymmetric Theories}
\label{radiative}

Supersymmetric theories are immunized against many of the types of corrections found in non-supersymmetric theories as a consequence of holomorphy and non-renormalizations.  In particular,
large terms of the form $H^* H \eta^*_i \eta_j$  and $\eta_i \eta_j \eta_k^* \eta_l^*$ need not arise (the corresponding superpotential
terms can be suppressed by symmetries and the smallness of the $\mu$ term).  There are, however, new possible sources of
corrections to $\theta$.  We divide our discussion between gravity mediated and gauge mediated models.
Loop corrections in gravity mediated models, as discussed in \cite{dinekaganleigh}, are quite problematic.
Gauge mediated models are better controlled~\cite{hillerschmaltz}.

We assume that tree level contributions to phases of gaugino masses are highly suppressed.  Beyond this, we require, as discussed
above, suppression of phases in the underlying supersymmetry breaking $f$ term and the superpotential.  But there are still potential
difficulties.  As discussed in \cite{dinekaganleigh}, already at one loop, there are contributions to gaugino masses arising
from loops involving heavy fields in the CP violating sector.  In the simplest model, the heavy field is a Dirac particle, of mass $m_D$,
consisting of a charge $1/3$ field,
\beq
\bar D = \sum B_f \bar d_f + \mu \bar q
\eeq
and a field of charge $-1/3$, $D=q$.  There is a soft breaking term,
\beq
{\cal L}_{q\bar D} = A_D m_D \bar D D\;.
\eeq
The gluino mass receives contributions proportional to $A_D^*$.
In general, there is no reason for the phase of $A$ to vanish; this requires a very specific alignment of expectation values and 
couplings.  It could arise in the presence of an $SU(4)$ symmetry acting on $\bar d$ and $\bar q$ -- something clearly not
present in this structure.  The phase must be smaller than $10^{-8}$ or so.  Similarly, there are potential contributions
proportional to $F_{\eta_a}$.  In supergravity models, these may naturally be suppressed 
by $(m_{3/2}/M_p)^{1/2}$, so they become problematic if the scale of supersymmetry breaking is greater than $10^4$ GeV or so.

As discussed in \cite{dinekaganleigh}, there are additional contributions arising from phases in soft scalar mass terms.
Suppressing these requires a remarkably high degree of degeneracy and proportionality.  Overall, then, there is a set of issues similar
to, but more severe than, the usual flavor problems of supergravity theories.

Gauge mediated models are characterized by features which ameliorate the problems noted above.\footnote{See also the discussion in~\cite{Barr:1996wx} for the possibility of suppression through alignment.}  First and foremost, new sources of flavor violation are absent, and $A$ terms are highly suppressed. 

In addition, insertions of $F_{\eta_a}$, which also enter in loop corrections to gaugino masses, are small if SUSY breaking does not couple to the $\eta_a$ at tree level. SUSY-breaking $F$-terms for the $\eta_a$ are generated radiatively from Kahler potential operators such as $Z^\dagger Z \eta_a^\dagger \eta_b/m_{CP}^2$, but in the minimal model they appear only at three loop order.  These statements need not hold in theories where 
messengers mix with other fields so as to gain large $A$ terms, or where there are ``$\mu$-terms" for some of the $\eta$ fields.

At higher loop order, complex A-terms and flavor-violating soft masses can be generated in gauge mediation. Such terms can give a weak upper bound on the hierarchy $F_Z/m_{CP}^2$. For example, in minimal gauge mediation, a Kahler potential operator of the form $Z^\dagger Z q \bar d_f \eta_a / m_{CP}^3$ is generated at 3-loop order from loops of the $\eta$ fields connected to ordinary gauge mediation loops. This operator provides a phase to the gluino mass in a manner similar to a complex A-term of the form $A_\gamma  \eta q \bar{d}$ (although the operator involves heavy fields and cannot be written as an A-term at the scale $m_{CP}$). Because of the high loop suppression, the bound from $\bar \theta$ is weak: $F_Z/m_{CP}^2\lesssim10^{-2}$. 

Furthermore, all non-minimal flavor violation among the light fields comes from the coupling $a_{af}\eta_a \bar d_f q$ and the mixing of light right-handed fields with $a_{af}\langle \eta_a\rangle \bar d_f$. If $\mu\ll a_{af}\langle \eta_a\rangle$, the light field is mostly $\bar q$, and the mixing is small. Since $\mu \gg a_{af}\langle \eta_a\rangle$ is in conflict with the large CKM phase, and there is no obvious reason for the scales to be coincident, contributions to $\bar \theta$ in gauge-mediated NB models can be even further suppressed by $\mu/m_{CP}$.

\section{Conclusions}
\label{conclusions}

We have argued that solving the strong CP problem is not necessarily an arena for model building cleverness; rather, ideally, the smallness of an inconsequential parameter should emerge as a consequence of features of a theory which explains a range of other phenomena.  No currently known model for solving strong CP is completely satisfactory from this point of view.

The shortcomings of the axion solution are well-known. Perhaps the most credible realization is in string theory, where plausible assumptions about moduli fixing may lead to a solution, albeit with a relatively high-scale axion.  

In the case of the Nelson-Barr solution, we have argued that non-supersymmetric models are at best very complicated, with intricate symmetries required to suppress higher-dimension operators. If these operators are simply suppressed by a low scale of CP violation, models without strong dynamics or supersymmetry require a degree of fine-tuning higher than if $\bar \theta$ were simply set to zero by hand. Furthermore, we have argued that dynamical models based on vevs for pseudo-Goldstones are nontrivial to construct. 
Loop corrections in generic non-SUSY models are even more problematic, making further demands on the theories.

Supersymmetric Nelson-Barr fares somewhat better.  Coincidences of scales are still required, but light scalars can be technically natural, and holomorphy greatly restricts the higher-dimension operators that can contribute to $\bar \theta$.  We described a specific structure in which the NB mechanism is operative and CP is broken in approximate flat directions by fields carrying new gauge symmetries.  Additional discrete symmetries can suppress dangerous couplings of the CP-violating fields to the hidden sector fields and also couplings to messengers.  Loop corrections are known to be highly problematic in generic gravity-mediated models, but in gauge-mediated models, these effects are under control. So supersymmetric models with additional symmetries and gauge mediation provide a setting in which the Nelson-Barr mechanism is plausible, at least as viewed at relatively low scales.  

We have also studied the underlying premise of models that aim to solve the strong CP problem through spontaneous CP violation:  that in such theories, the bare $\theta$ parameter naturally vanishes.  We stressed that this is a question of the nature of the ultraviolet theory.  In string theory, the value of $\theta$ is generally controlled by the value of an axion field, so the basic assumption is that there are massive axions whose expectation values conserve CP.  Perhaps most problematic for the idea of small $\theta$, however, is the possibility of a landscape.  We noted that in flux landscapes, in particular, where the heavy axion expectation value is determined by superpotential parameters, these parameters are likely to be complex in an overwhelming majority of states.  

So the current status of the strong CP problem can be described by saying we possess three solutions, each with significant flaws.  The reader is free to develop his or her own view as to which solution, is any, is most plausible.  Unless there are systematic problems with lattice computations which are common to disparate approaches to QCD, the light $u$ quark solution is ruled out.  The axion solution requires either very complicated symmetry structures, or some assumptions about moduli stabilization and an unconventional cosmological history.  The spontaneous CP solution requires supersymmetry, a variety of additional symmetries, something like gauge mediation, and, perhaps most problematic, an explanation of why moduli are stabilized in a CP-conserving way.

\vspace{1cm}

\noindent
{\bf Acknowledgements:}  This work was supported by the U.S. Department of Energy grant number DE-FG02-04ER41286.  We appreciate conversations with Nima Arkani-Hamed, Nathaniel Craig, Ravi Kuchimanci, Anson Hook, Nathan Seiberg, Goran Senjanovic, Scott Thomas, and Luca Vecchi. 
\bibliography{dinerefs}
\bibliographystyle{jhep}

\end{document}